\documentclass{article}
\usepackage{amssymb}

\def\R{{\mathbb R}}

\begin{document}

\title{Construction of oscillatory singularities} 

\date{}

\maketitle

\begin{abstract}
One way to understand more about spacetime singularities is to construct
solutions of the Einstein equations containing singularities with prescribed
properties. The heuristic ideas of the BKL picture suggest that oscillatory
singularities should be very common and give a detailed picture of how these
could look. The more straightforward case of singularities without 
oscillations is reviewed and existing results on that subject are surveyed. 
Then recent theorems proving the existence of spatially homogeneous solutions 
with oscillatory singularities of a specific type are presented. The proofs of 
these involve applications of some ideas concerning heteroclinic chains and 
their stability. Some necessary background from the theory of dynamical 
systems is explained. Finally some directions in which this research might be 
generalized in the future are pointed out.
\end{abstract}

\section{Introduction}

One of the characteristic features of general relativity is its prediction of
spacetime singularities. These can already be observed in explicit solutions
of the Einstein equations such as the Schwarzschild solution and the 
Friedman-Lemaitre-Robertson-Walker (FLRW) solutions. These explicit solutions
have a high degree of symmetry. The question arose early whether the known 
singularities might be artefacts of symmetry. The work of Lifshitz and
Khalatnikov \cite{lifshitz} supported this idea. These authors made what
they believed to be a general ansatz for the form of the geometry near the
singularity and found that it could not accomodate the full number of free
functions expected. They concluded that generic solutions of the Einstein
equations do not develop singularities. The result was heuristic in nature.

The conclusions of \cite{lifshitz} were shown to be invalid by the 
singularity theorems of Penrose and Hawking \cite{hawking}. In contrast to
the arguments in \cite{lifshitz} the results of the singularity theorems were 
based on mathematical proofs. It was shown that singularities occur for open 
sets of solutions of the Einstein equations. More specifically, using the 
point of view of the initial value problem, they occur for the solutions arising
from open sets of initial data. This means that the occurrence of singularities
is stable in a certain precise sense. A positive feature of these theorems is 
that the hypotheses are relatively weak. On the other hand the conclusions are 
also relatively weak. What is proved is that the spacetimes to which the 
theorems apply are geodesically incomplete. Very little is said about the 
nature of the singularities. There is no information on whether the 
singularities are accompanied by large energy densities or tidal forces, as 
might be expected on the basis of physical intuition. 

Later, Belinskii, Khalatnikov and Lifshitz (abbreviated in what follows by BKL)
presented a more detailed picture of spacetime singularities 
\cite{belinskii70}. Their arguments are heuristic. They are similar to the 
work of \cite{lifshitz} with the important difference that this time the ansatz
used is more complicated, including oscillatory behaviour. This allows the 
full number of free functions to be included. Some of the main assertions
belonging to the BKL picture are:

\begin{itemize}
\item the solutions of the partial differential equations are approximated
by solutions of ordinary differential equations near the singularity
\item the approach to the singularity is oscillatory
\item solutions of the Einstein equations with matter are approximated by 
solutions of the vacuum equations near the singularity
\end{itemize}    

\vskip 10pt
The solutions of the ordinary differential equations in the first point 
correspond to spatially homogeneous solutions of the Einstein equations. In 
the original work one of the
most general classes of spatially homogeneous solutions, the Bianchi type IX
solutions, played a central role. These solutions were also studied 
independently at about the same time by Misner \cite{misner} who called them 
the Mixmaster model. On the basis of heuristic and numerical work the conclusion
was reached that these solutions show a highly oscillatory behaviour near the
singularity. This statement is correct but, as will be discussed below, it 
took a long time to prove it. Here the connection to the second point above
can be seen. Moreover, the oscillations are observed in the vacuum case and
this makes contact with the third point. From these remarks it follows that
if the BKL picture is correct then spatially homogeneous solutions of the 
vacuum equations are very important in understanding singularities in 
solutions of the Einstein equations without symmetries and with quite general
matter. Note, however, that after forty years it is still not known whether
the BKL picture is correct. At least there have been many studies of 
solutions with symmetries which give good agreement with the conclusions
obtained by specializing the general BKL picture to symmetric cases. 

Since the general case is very complicated to treat it makes sense to start 
looking for a better understanding of the question of the validity of the BKL
picture by concentrating on the spatially homogeneous case. This is especially 
true due to the fact that the wider significance of the homogeneous case is 
intimately related to the main claims of the BKL picture. This leads to the
study of certain systems of ordinary differential equations. A concept from
the theory of ODE which turns out to be of particular relevance in this 
context is that of heteroclinic chains. They will be discussed in Sect.
\ref{heteroclinic}. In special cases the dynamics near the singularity is 
convergent rather than oscillatory and in this easier case there do exist 
results for inhomogeneous spacetimes. Since these are relevant for the 
conceptual approach used in the case of oscillatory asymptotics they will be 
discussed in Sect. \ref{convergent}. In order to even define the concepts 
\lq monotone\rq\ and \lq oscillatory\rq\ being used it is necessary to 
introduce some basic notation and terminology and this is the subject of 
Sect. \ref{notation}. 

\section{Notation and terminology}\label{notation}

Let $M$ be a four-dimensional manifold and $g_{\alpha\beta}$ a Lorentzian 
metric on $M$. Let $S_t$ be a foliation by spacelike hypersurfaces whose 
leaves are parametrized by $t$. Let $g_{ab}(t)$ and $k_{ab}(t)$ be the induced 
metric and the second fundamental form of the leaves of the foliation. Define
the Hubble parameter as $H=-\frac13 g^{ab}k_{ab}$. It is possible to do a 
3+1 decomposition of the Einstein equations for the metric 
$g_{\alpha\beta}$ based on the foliation $S_t$ which brings in a lapse function 
and a shift vector. This can of course be done in the presence of matter and 
also in a closely analogous way in other dimensions.

Let $\lambda_i$ be the eigenvalues of $k_{ab}$ with respect to $g_{ab}$, i.e.
the solutions of $k_{ab}X^b=\lambda_i g_{ab}X^b$ for some vector $X^b$. For a
well-behaved foliation approaching the singularity towards the past $H>0$ 
in a neighbourhood of the singularity. Hence the quantities 
$p_i=-\lambda_i/3H$, the generalized Kasner exponents, are well-defined 
functions of $t$ and $x$, where $x$ denotes a point on one of the leaves
of the foliation. It follows from the definition that $\sum_i p_i=1$ so
that only two of the $p_i$ are independent.
The way in which different leaves are identified with each
other depends on the choice of the shift vector. Suppose that the limit
$t\to 0$ corresponds to the approach to a singularity. Fix a point $x_0$ and 
consider the functions $p_i(t,x_0)$. A function $p_i$ is said to be
oscillatory near the singularity if 
$\liminf_{t\to 0} p_i(t)<\limsup_{t\to 0} p_i(t)$. It is said to be convergent
near the singularity if $\liminf_{t\to 0} p_i(t)=\limsup_{t\to 0} p_i(t)$, in
other words if $p_i$ tends to a limit as $t\to 0$. Similar definitions can
be made for other geometric quantities. Informally, a singularity is said to
be oscillatory or convergent if some important geometric quantities have
the corresponding properties. It is important that the quantities concerned
are dimensionless, since otherwise they could be expected to diverge as the
singularity is approached. 

These definitions depend a lot on the choice of 3+1 decomposition. Things are
simpler in the case of spatially homogeneous spacetimes. There it is natural
to choose the foliation to consist of level hypersurfaces of a Gaussian 
time coordinate based on a hypersurface of homogeneity. Then geometric 
quantities such as $p_i$ depend only on $t$.  

\section{The convergent case}\label{convergent}

The second and third points listed above as parts of the BKL picture are only 
supposed to hold in this form with certain restrictions on the type of 
matter which is coupled to the Einstein equations. Assuming the validity of 
the first point this is directly related to the behaviour of homogeneous 
models with that type of matter. Two related types of matter which are 
known to be exceptional from this point of view are the scalar field and
the stiff fluid. Here we will concentrate on the second. Consider a 
perfect fluid with linear equation of state $p=(\gamma-1)\rho$ for a 
constant $\gamma$. An extreme case of this from the point of view of 
ordinary physics is $\gamma=2$. This is the stiff fluid where the velocity of 
sound of the fluid $\sqrt{\gamma-1}$ is equal to the velocity of light. (We use
geometrical units.) It is also possible to consider ultrastiff fluids with
$\gamma >2$. These very exotic matter models have been considered in certain
scenarios for the early universe. According to the BKL picture the generalized 
Kasner exponents in a general spatially homogeneous solution of the 
Einstein-Euler equations with a linear equation of state are oscillatory near 
the singularity in the case $1\le\gamma<2$ and convergent for $2\le\gamma$. 
(The case $\gamma<1$ will not be considered in what follows.) In fact for 
$\gamma>2$ the $p_i$ all converge to the limit $\frac13$. These limiting values
agree with the values of these quantities in the FLRW solutions and 
correspond to isotropization. In the case $\gamma=2$, on the other hand, the 
$p_i$ may converge to different limits in different solutions and at different 
spatial points in a single inhomogeneous solution.

What kind of mathematical results could count as a rigorous version of these
expectations? It would be desirable to prove that an open neighbourhood with
respect to some reasonable topology of the FLRW data leads to solutions which
have the predicted asymptotics. This may be called the forwards result since
it goes from the data to the asymptotics. Unfortunately this kind of result 
is hard to obtain. What is sometimes easier is to get a backwards result
which goes from the asymptotics to the solution. The solutions with the 
predicted asymptotics can be parametrized by certain free functions. (Here we
ignore complications arising from the Einstein constraints.) Call these 
functions the asymptotic data. The idea is then to ask whether for  
asymptotic data belonging to a large class there exists a solution with the 
correct asymptotics and exactly those free functions. A theorem of this kind 
was proved for stiff fluids in \cite{andersson01a} and for ultrastiff fluids 
in \cite{heinzle12}. In this work the notion of \lq large 
class\rq was defined to mean containing as many free functions as the general
solution. It was only proved in the case that these functions are analytic
($C^\omega$). Until very recently the forwards problem was unsolved, even in 
the stiff case, but a theorem of this kind has been announced by Rodnianski 
and Speck \cite{rodnianski12}.   

For comparison it is of interest to quote what is known in a case with symmetry,
that of the Gowdy solutions. Note that for the Gowdy solutions the BKL
picture does not predict oscillations and indeed none are found. In this
case the backwards problem was solved in \cite{kichenassamy98} for the analytic
case and in \cite{rendall00a} for the smooth case. The forwards problem for
Gowdy was solved in \cite{ringstrom04}. The aim of this section was to exhibit 
the backwards problem as a route to obtaining rigorous results and to show that
at least in some cases it can open the way to obtaining results of the type 
which are most desirable.

Just as some matter models such as the stiff fluid can suppress oscillations
others can stimulate them. For instance a Maxwell field can produce BKL type
oscillations within Bianchi types where there are no oscillations in the 
vacuum case. An example is discussed in Sect. \ref{construction}. The BKL 
picture can be applied in higher dimensions and gives different results in 
that case. It says, for instance, that in the case of the Einstein vacuum
equations generic oscillations disappear when the spacetime dimension is 
at least eleven but are present in all lower dimensions. The higher dimensional
models come up in the context of string theory and there it is typical that 
there are other matter fields present. There is the dilaton which, as a scalar
field, tends to suppress oscillations and the $p$-forms which, as 
generalizations of the Maxwell field, encourage them. Many of the conclusions
of the heuristic analysis which lead to the conclusion that the singularity
is convergent can be made rigorous using techniques generalizing those of 
\cite{andersson01a}. For more information on this the reader is referred to
\cite{damour02}. To conclude it should be noted that there is not a single 
rigorous result which treats solutions which are both inhomogeneous and 
oscillatory.

\section{Heteroclinic chains}\label{heteroclinic}

Consider a system of $k$ ordinary differential equations. This can be thought
of as being defined by a vector field defined on an open subset of $\R^k$, 
a geometrical formulation which corresponds to the point of view of 
dynamical systems. A stationary (i.e. time-independent) solution of the 
ODE system corresponds to a fixed point of the vector field. A solution
of the ODE system which is time-dependent corresponds to an integral 
curve of the vector field, an orbit of the dynamical system. An orbit 
which tends to a stationary solution $p$ as $t\to -\infty$ and a stationary
point $q\ne p$ as $t\to +\infty$ is called a heteroclinic orbit. An orbit
which tends to $p$ as $t\to -\infty$ and as $t\to +\infty$, but is not 
itself stationary, is called a homoclinic orbit.

Suppose now that $p_i$ is a sequence of points, finite or infinite, such
that for each $i$ there is a heteroclinic orbit from $p_i$ to $p_{i+1}$.
This configuration is called a heteroclinic chain. If the chain is finite
and comes back to its starting point then it is called a heteroclinic cycle.
It turns out that homoclinic orbits and heteroclinic cycles are not
robust in the sense that a sufficiently general small perturbation of the
vector field will destroy them. It is therefore perhaps surprising that
they are found in models for many phenomena in nature. Given that all
physical measurements only have finite precision it seems difficult
to be able to rule out small perturbations in mathematical models for
real phenomena. The answer to this question appears to be that there is
some absolute element in the system whose presence is not subject to 
uncertainty and which therefore cannot be perturbed. In the case of 
spacetime singularities it is the singularity itself which appears 
to play this role. These considerations are admittedly rather vague and
it would be desirable to understand issues of this type more precisely. In 
any case, heteroclinic cycles do seem to be widespread in models for the
dynamics of solutions of the Einstein equations near their singularities.
 
Coming back to more general dynamical systems, it is of interest to have
criteria for the stability properties of heteroclinic chains. In other
words, the aim is to find out under what conditions solutions which 
start close to a heteroclinic chain approach it as $t\to\infty$. The 
case of most interest is not that where a solution approaches one of 
the stationary points belonging to the heteroclinic chain (call them 
vertices) but the case where the solution follows successive heteroclinic
orbits within the chain. When the chain is a cycle this means that the
solutions exhibit oscillatory behaviour, repeatedly approaching the 
vertices of the cycle. In the next section it is shown that this is 
exactly what happens in the context of the BKL picture.

A solution which is converging to a heteroclinic chain spends most of its time
near the vertices. Since the vector field vanishes at the vertices it is
small near them and the solution moves slowly there. This suggests that the
behaviour of the solution while it is near the vertices could have strong 
influence on the stability of the cycle. A way of studying the local flow near
a vertex is to linearize there and examine the eigenvalues of the 
linearization. It turns out that this can give valuable information about 
the stability of the cycle.

\section{Bianchi models}\label{bianchi}

A spatially homogeneous solution of the Einstein equations is one which
admits a symmetry group whose orbits are spacelike hypersurfaces. The only
case where the group cannot be assumed to be three-dimensional is the class
of Kantowski-Sachs spacetimes. They are not discussed further in this paper.
All the rest are the Bianchi models. In that case it can be assumed without
loss of generality that the spatial manifold is simply connected and then it 
can be identified with the Lie group itself. The reason for this is that the 
dynamics on the universal covering manifold is the same as that on the original 
manifold. Thus in Bianchi models it can be assumed that $M=I\times G$ where
$I$ is an interval and $G$ is a simply connected three-dimensional Lie group.
These Lie groups are in one to one correspondence with their Lie algebras
and the three-dimensional Lie algebras were classified by Bianchi into types
I-IX. The metric can be written in the form
\begin{equation}
-dt^2+\sum g_{ij}(t)\theta^i\otimes \theta^j
\end{equation}
where $\theta^i$ is a basis of left invariant one-forms on $G$. Let $e_i$ be
the dual basis of vector fields. The evolution equations depend on the Lie 
group chosen through the structure constants defined by 
$[e_i,e_j]=c^k_{ij}e_k$. It is customary to distinguish between Class A models
where $c^k_{kj}=0$ and Class B models which are the rest.

There are many ways in which the evolution equations for Bianchi models can be
written as a system of ordinary differential equations. Consider for simplicity
the vacuum models of Class A. In that case it can be shown that there is a 
basis of the Lie algebra with the property that diagonal initial data give
rise to diagonal solutions. This is still true when the matter is described
by a perfect fluid but it does not hold for general matter. A form of the 
equations for Class A models which has turned out to be particularly useful
for proving theorems is the Wainwright-Hsu system \cite{wainwright89}. The 
basic variables are called $\Sigma_+,\Sigma_-,N_1,N_2,N_3$. The first two are
certain linear combinations of the generalized Kasner exponents. The 
quantity $N_1$ is given in the Bianchi type IX case by 
$\frac{1}{H}\sqrt\frac{g_{11}}{g_{22}g_{33}}$ 
and the other two $N_i$ are related to this by cyclic permutations. The Gaussian
time coordinate $t$ is replaced by a coordinate $\tau$ which satisfies
$\frac{d\tau}{dt}=H$. In this time coordinate the singularity is approached
for $\tau\to -\infty$. Two important related features of this system is that 
the variables are dimensionless and that their evolution equations form a 
closed system, not depending on $H$. They must satisfy one algebraic condition
which is the Hamiltonian constraint in the 3+1 formalism. There results a 
four-dimensional dynamical system. Stationary solutions of this system 
correspond to self-similar spacetimes. When a solution of the Wainwright-Hsu
system is given the quantities $H$ and $t$ can be determined by integration. 

The different Bianchi types are represented by subsets of the state space of 
the Wainwright-Hsu system where the $N_i$ have certain combinations of signs 
(positive, negative or zero). The simplest Bianchi type is Bianchi type I, 
the Abelian Lie algebra. The vacuum solutions of this type are the Kasner 
solutions. They are self-similar and form a circle in the state space, the 
Kasner circle $\cal K$. Let $\cal T$ be the equilateral triangle 
circumscribing the 
Kasner circle which is symmetric under reversal of $\Sigma_-$. It is tangential 
to $\cal K$ at three points $T_i$, the Taub points. They correspond to flat 
spacetimes. Each solution of Bianchi type II is a heteroclinic orbit joining 
two Kasner solutions. Its projection to the $(\Sigma_+,\Sigma_-)$-plane is a 
straight line which passes through a corner of $\cal T$. Concatenating Bianchi 
type II solutions gives rise to many heteroclinic chains. Given a point of
$\cal K$ which is not one of the Taub points consider the straight line
joining it to the closest corner of the triangle $\cal T$. The part of this
straight line inside $\cal K$ is the projection of an orbit of type II and 
it intersects $\cal K$ in exactly one other point. In this way it is possible 
to define a continuous map from $\cal K$ to itself, the BKL map. (The map is 
defined to be the identity at the Taub points.) The two most general Bianchi 
types of Class A are type VIII and type IX, corresponding to the Lie algebras 
$sl(2,\R)$ and $su(2)$. As already mentioned BKL concentrated on type IX. Type 
VIII shows many of the same features but may be even more complicated.

When specialized to the Bianchi Class A case the BKL picture suggests that 
as the singularity is approached in a Bianchi type IX solution the dynamics
should be approximated by a heteroclinic chain of Bianchi type II solutions,
successive vertices of which are generated by the BKL map. In particular the 
solution is oscillatory in the sense that the $\alpha$-limit set of a solution 
of this kind should consist of more than one point. (The $\alpha$-limit set 
consists of those points $x$ such there is a sequence of times tending to 
$-\infty$ along which the value of the solution converges to $x$.) In fact 
there should be three non-collinear points in the $\alpha$-limit set. This 
implies that both $\Sigma_+$ and $\Sigma_-$ are oscillatory in the approach to 
the singularity. This statement about the $\alpha$-limit set was proved in 
fundamental work of Ringstr\"om \cite{ringstrom00}. He showed that there are 
at least three non-Taub points in the $\alpha$-limit set of a generic Bianchi IX
solution. (He also described the exceptions explicitly.) In addition 
Ringstr\"om showed that the entire $\alpha$-limit set of a Bianchi type IX 
solution is contained in the union of the points of type I and II 
\cite{ringstrom01}. This means that in some sense the type IX solution 
is approximated by solutions of types I and II. The question of whether 
the corresponding statement holds for solutions of Bianchi type VIII is still
open.

After the results just discussed it still took a long time before theorems 
about convergence to heteroclinic cycles were published. There is a 
heteroclinic cycle of Bianchi II solutions which comes back to its starting
point after three steps. Let us call this particularly simple example \lq the
triangle\rq. It turns out that there is a codimension one manifold with the 
property that any Bianchi type IX solution which starts on this manifold 
converges to the triangle \cite{liebscher11} in the past time direction. This 
result extends to a much larger class of heteroclinic chains generated by 
iterating the BKL map. The essential condition is that the vertices of the 
chain should remain outside an open neighborhood of the Taub points. The 
manifold constructed in \cite{liebscher11}, which may referred to as the 
unstable manifold of the triangle (unstable towards the future and hence stable
towards the past), is only proved to be Lipschitz continuous. An alternative 
approach to this problem was given in \cite{beguin10}. It has the advantage 
that the stable manifold is shown to be continuously differentiable. On the 
other hand it cannot treat all the heteroclinic chains covered by the results
of \cite{liebscher11}. In particular, it does not cover heteroclinic cycles
such as the triangle. The reason for this restriction is the need to 
avoid resonances, certain linear relations between the eigenvalues with integer 
coefficients. When this extra condition is satisfied it can be shown, using 
a theorem of Takens, that the flow near any vertex of the chain is 
equivalent to the linearized flow by a diffeomorphism. Some investigations
of the case of chains which may approach the Taub points have been carried 
out in \cite{reiterer10}.  

\section{Construction of solutions converging to the triangle}
\label{construction}

The construction of solutions converging to heteroclinic chains of Bianchi type 
II solutions in the approach to the initial singularity will be illustrated 
by the case of the triangle. As indicated in Sect.  \ref{heteroclinic}
the stability of heteroclinic chains is related to the eigenvalues of the 
linearization about the vertices. Since a vertex of the triangle lies on
the Kasner circle which consists of stationary points the linearization
automatically has a zero eigenvalue. Of the three other eigenvalues one
is negative, call it $-\mu$, and two are positive, call them $\lambda_1$
and $\lambda_2$. The two positive eigenvalues are distinct and we adopt the 
convention that $\lambda_1<\lambda_2$. The eigendirection corresponding to
$\lambda_2$ is tangent to the triangle. The proofs of \cite{liebscher11} are
dependent on the fact that $\mu$ is smaller than both positive eigenvalues.

A solution which converges to the triangle repeatedly passes the vertices.
Consider a point on the triangle which is close to a vertex and on an orbit
approaching it. Let $S$ be a manifold passing through the point which is 
transverse to the heteroclinic orbit. Call this an incoming section. Similarly 
we can consider an outgoing section close to the vertex and passing through 
the orbit leaving it. A solution which starts on the incoming section 
sufficiently close to the heteroclinic cycle also intersects the outgoing 
section. Taking the first point of intersection defines a local mapping 
from the incoming section to the outgoing section, the local passage.
There is a similar local mapping from the outgoing section of one vertex to
the ingoing section of the next in the cycle. This is called an excursion.
Composing three passages and three excursions gives a mapping from the
ingoing section of a vertex to itself. It is important for the proof of
\cite{liebscher11} that this mapping is Lipschitz and that by restricting
the domain of the mapping to a small enough neighbourhood of the triangle
the Lipschitz constant can be made as small as desired. The norm used to
define the Lipschitz property is that determined by the flat metric 
$dx^2+dy^2+dz^2$, where $(x,y,z)$ are some regular coordinates on the 
section. The mapping from a small local section to itself is a 
contraction. The excursions are expanding mappings but the expansion
factor is bounded. The passages are contractions which can be made 
arbitrarily strong and which can therefore dominate the effect of the 
excursions. Once the contraction has been obtained the manifold
being sought can be constructed in a similar way to the stable manifold 
of a stationary solution. 
 
A system similar to the Wainwright-Hsu system can be obtained for solutions 
of Bianchi type VI${}_0$ with a magnetic field \cite{leblanc95}. It also 
includes certain solutions of Bianchi types I and II with magnetic fields and 
vacuum solutions of types I, II and VI${}_0$. It uses variables $\Sigma_+$ and
$\Sigma_-$ which have the same geometrical interpretation as in the vacuum
case. The Kasner circle can be considered as a subset of the state space
for the magnetic system. It has been proved that solutions of this system
are oscillatory \cite{weaver}. There are two families of heteroclinic orbits
defined by vacuum solutions of type II. The third family in the vacuum 
case is replaced by heteroclinic orbits defined by Bianchi type I solutions
with magnetic fields. The projections of these heteroclinic orbits onto
the $(\Sigma_+,\Sigma_-)$-plane are the same straight lines as are obtained
from heteroclinic orbits in the vacuum case. Thus the exactly the same 
heteroclinic chains are present. However their stability properties might 
be different. Now the stability of the triangle will be considered in the
case with magnetic field, following \cite{liebscher12}. The result is very 
similar to that in the vacuum case - there is a one-dimensional unstable 
manifold - but the proof is a lot more subtle. The reason that the method of
proof of \cite{liebscher11} does not apply directly is that the eigenvalue 
configuration is different. Compared to the vacuum case one of the eigenvalues
is halved. Then it can happen that the negative eigenvalue is larger in modulus
than one of the positive eigenvalues.

A problem which results from the different eigenvalue configuration is that 
the return map is no longer Lipschitz. More precisely it no longer has this
property with respect to the Euclidean metric $dx^2+dy^2+dz^2$. It does,
however, have the desired Lipschitz property with respect to the 
singular metric 
$\frac{x^2+y^2}{x^2}dx^2+\frac{x^2+y^2}{y^2}dy^2+dz^2$ 
and this observation allows the proof of \cite{liebscher11}
to be generalized to the case with magnetic field. It is important to know 
that the properties of eigenvalues and eigenvectors alone are not enough
to make this proof work. It is also necessary to use the existence of 
certain invariant manifolds which follows form the geometric background of
the problem. At this point it is necessary to remember that we are not just
dealing with a heteroclinic cycle in an arbitrary dynamical system but with
one in a system with very special properties.

It turns out that the difficulties just discussed can be avoided by a clever
but elementary device which comes down to replacing the variable representing
the magnetic field by its square. Although this provides a very simple way of
studying the heteroclinic chains in the Bianchi VI${}_0$ model with magnetic 
field it cannot be expected that this kind of trick will apply to more general 
matter models. By contrast the new method is potentially much more generally
applicable. There is one case where it is already known to give new results,
as will now be explained.
In addition to the results obtained on vacuum models results on models with
a perfect fluid with linear equation of state $p=(\gamma -1)\rho$ were 
obtained in \cite{liebscher11}. It turns out, however, that the techniques
of \cite{liebscher11} only work under an assumption on $\gamma$ which has
no physical interpretation. In the case of the triangle the condition is
$\gamma<\frac{5-\sqrt{5}}{2}\sim 1.38$. For other chains other inequalities
are obtained. These arise because the linearization has another positive 
eigenvalue coming from the fluid and it must be ensured that this eigenvalue
is greater than $\mu$. With the method of \cite{liebscher12} this restriction
can be replaced by the inequality $\gamma<2$, saying that the speed of sound
in the fluid is smaller than the speed of light.  
 
\section{Future challenges}

This section discusses some directions in which the known results on the 
construction of oscillatory singularities might be extended in the future. There
are dynamical systems describing Bianchi models of types I and II with 
magnetic fields which are not just special cases of those included in the 
system describing models of type VI${}_0$. The reason for this is that the 
Maxwell constraints become less restrictive in the more special Bianchi types.
It is possible to choose the basis so that the magnetic field only has one 
non-vanishing component but the price to be payed is that the metric becomes
non-diagonal in that basis. In the dynamical systems describing solutions of
types I and II the already familiar heteroclinic chains are still found.
There are, however, additional heteroclinic orbits corresponding to the 
off-diagonal elements of the metric. This means that, in contrast to the 
models analysed up to now, the stable manifold of a point on the Kasner circle
may be of dimension greater than one. This means that when it is desired to 
continue a heteroclinic chain towards the past there is not a unique choice
any more. New ideas are required to solve this type of problem.

Similar difficulties are met in Bianchi models of Class B. There it is 
believed on the basis of heuristic considerations that there is precisely
one type, VI${}_{-\frac19}$, which shows oscillatory behaviour similar to that 
found in types VIII and IX. There are no rigorous results on Class B
comparable to the results of Ringstr\"om on models of Class A. In this case 
too the stable manifold of a Kasner solution may have dimension greater than 
one. It can also not be expected that invariant manifolds of the type 
exploited in \cite{liebscher12} will exist.

Perhaps the most exciting challenge in this field is to construct inhomogeneous 
spacetimes with oscillatory singularities. The simplest class of
inhomogeneous vacuum spacetimes where oscillations are expected are the 
$T^2$ models. These have a two-dimensional isometry group acting on 
spacelike hypersurfaces and so are effectively inhomogeneous in just one
space dimension. They include the Gowdy spacetimes as a subset but it
is believed that generic $T^2$-symmetric vacuum spacetimes have a much 
more complicated oscillatory behaviour near the singularity. The models
of Bianchi type VI${}_{-\frac19}$ are locally isometric to 
$T^2$ models but not locally isometric to Gowdy models. Thus understanding
more about Bianchi models of Class B appears a very natural first step 
towards a better understanding of the inhomogeneous case.

\end{document}